\documentclass[runningheads]{llncs}
\usepackage[T1]{fontenc}
\usepackage{ragged2e}
\usepackage{graphicx}
\usepackage{amsmath}
\usepackage{physics}
\usepackage{braket}
\usepackage{comment}
\usepackage{hyperref}
\usepackage{color}

\usepackage{xurl}
\begin{document}
\title{Understanding Quantum Imaginary Time Evolution and its Variational form}
\titlerunning{Understanding QITE and varQITE}

\author{Andreu Anglés-Castillo\inst{1}\orcidID{0000-0003-2883-4851}\email{aangcas@upv.es} \and
Luca Ion\inst{2} \and
Tanmoy Pandit\inst{3,4,5}\orcidID{0000-0002-2377-6181} \and 
Rafael Gomez-Lurbe \inst{2} \and
Rodrigo Martínez\inst{6} \and
Miguel Angel Garcia-March \inst{1}
}
\authorrunning{A. Anglés-Castillo et al.}

\institute{Instituto Universitario de Investigación de Matemática Pura y Aplicada, Universitat Politècnica de València, València, Spain \and Departament de Física Teòrica, Universitat de València, València, Spain \and Fritz Haber Research Center for Molecular Dynamics, Hebrew University
of Jerusalem, Jerusalem 9190401, Israel\and Institute for Theoretical Physics, Leibniz Institute of Hannover, Hannover, Germany\and 
 Institute for Theoretical Physics, TU Berlin, Berlin, Germany \and
Departament d’Informàtica de Sistemes i Computadors, Universitat Politècnica de València, València, Spain  }

\maketitle              
\begin{abstract}
Many computationally hard problems can be encoded in quantum Hamiltonians. The solution to these problems is given by the ground states of these Hamiltonians. A state-of-the-art algorithm for finding the ground state of a Hamiltonian is the so-called Quantum Imaginary Time Evolution (QITE) which approximates imaginary time evolution by a unitary evolution that can be implemented in quantum hardware. In this paper, we review the original algorithm together with a comprehensive computer program, as well as, the variational version of it.

\end{abstract}
\section{Introduction}

Quantum technologies are a promising avenue for solving many hard problems that are intractable with current computing techniques. A broad class of NP-problems can be encoded into Ising type Hamiltonians \cite{Lucas2014}
by mapping the solutions to these problems to the ground state of a particular Ising Hamiltonian. It is also possible to encode Boolean and real functions into Hamiltonian systems \cite{Hadfield2021}. Moreover, knowing the ground state is key for studying the quantum phases of matter, understanding material properties, and describing the chemical properties of different molecules

The canonical approach for obtaining the ground state of a Hamiltonian is to diagonalize it. This can always be done for Hermitian matrices; the diagonalization is understood as a diagonal block matrix for the case where there are degenerate eigenvalues. The density matrix renormalization group (DMRG) \cite{NAKATANI2018} is a classical variational algorithm for resolving low-energy physics of many body systems that has proved successful for solving 1D systems and extensions for higher dimensional systems exist \cite{RevModPhys.93.045003}.

In the domain of quantum many-body , hybrid quantum-classical algorithms such as the Variational Quantum Eigensolver (VQE) \cite{peruzzo2014variational} and the Quantum Approximate Optimization Algorithm (QAOA) \cite{farhi2014quantum} have been proposed to approximate ground states using parameterized quantum circuits. VQE is particularly useful for simulating molecular systems and quantum materials, while QAOA is tailored to solve combinatorial optimization problems encoded in Ising-type Hamiltonians. These algorithms are designed to work on near-term noisy quantum hardware, offering a scalable alternative to exact diagonalization. Also, let us remark on the quantum adiabatic computing (QAC) approach to find ground states, which is proven to be equivalent to quantum gate computing~\cite{adiabatic}. In the QAC approach,  an initial easy Hamiltonian, with a known ground state, is transformed into a final Hamiltonian that encodes the solution of the problem at hand. 

Recently, a new proposal for finding the ground state of Hamiltonian systems appeared \cite{Motta2019DeterminingEA}, which approximates an Imaginary Time Evolution (ITE) with unitary operators. This method is deterministic and avoids some of the convergence problems of variational algorithms. Then, in section 2 of this paper we give a detailed description of it, while in section 3 we formulate a variational alternative introduced in \cite{mcardle2019variational}. Finally, in section 4 we showcase both methods for a widely used model and discuss the tradeoffs and benefits of each method.

\section{Quantum Imaginary Time Evolution}

Obtaining the ground state $\ket{\psi_\mathrm{gs}}$ of a Hamiltonian $H$ for an $N$-spin system is an exponentially hard task as the system size increases. One way to obtain the ground state is performing imaginary time evolution (ITE) on a state $\ket{\psi_0}$ that has overlapping support with the ground state, i.e.,
\begin{equation}\label{eq:overlap}
	\Braket{\psi_0|\psi_\mathrm{gs}} \neq 0~.
\end{equation}
The ITE is performed as
\begin{equation} \label{eq:ITE}
	\ket{\psi(\beta)} = \frac{e^{-\beta H}\ket{\psi_0}}{\sqrt{\Braket{\psi_0|e^{-2\beta H}|\psi_0}}}~,
\end{equation}
whence the name of the method is derived, i.e., evolving the initial state in imaginary time $\beta = i t$. This dynamical state asymptotically approaches the ground state iff condition \eqref{eq:overlap} is fulfilled, i.e., $\lim_{\beta \to \infty} \ket{\psi(\beta)} = \ket{\psi_{gs}}$.
This unitary evolution is non-linear due to the state-dependent normalizations for each $\beta$. The Quantum ITE (QITE) method introduced in \cite{Motta2019DeterminingEA} consists of dividing the evolution in discrete time steps $\Delta \tau$ in imaginary time and finding unitaries that approximate the evolution of ITE, that is
\begin{equation}
	\ket{\psi_m} = \prod_{n=0}^m \frac{1}{c_n}e^{-n \Delta \tau H}\ket{\psi_0} \approx \prod_{n=0}^m e^{-i \Delta \tau A_n}\ket{\psi_0}\equiv \ket{\phi_m} ~,
\end{equation}
where $c_n=\sqrt{\Braket{\psi_0|\prod_{k=0}^n e^{-2 k \Delta \tau H}|\psi_0}}=\sqrt{\Braket{\psi_{n-1}|e^{-2 \Delta \tau H}|\psi_{n-1}}}$ is the normalization constant. We note that $\ket{\psi_m}$ and $\ket{\phi_m}$ differentiate the evolved state with ITE and QITE, respectively.
If such unitaries are found, the evolution can be carried out by quantum operations or, equivalently, in a quantum computer. The challenge of this algorithm lies in finding the unitaries that approximate the ITE. 
These unitaries are found after each time step
\begin{equation}
	\ket{\phi_{m}}=e^{-i \Delta \tau A_{m}}\ket{\phi_{m-1}}\approx \frac{1}{c} e^{-\Delta \tau H}\ket{\phi_{m-1}}~,
\end{equation}
where $c=\sqrt{\Braket{\phi_{m-1}|e^{-2 \Delta \tau H}|\phi_{m-1}}}$. Note here that we are using the state evolved with unitary evolutions $\ket{\phi_m}$ to compute an imaginary time step and finding the unitary that approximates that operation.

To find the unitary operator, we have to find the Hermitian operator $A_m$ at each time step. For qubit systems of $N$ sites, we can write $A_m$ as a sum of Pauli strings with real coefficients
\begin{equation}
	A_m = \sum_{i_1,i_2,...,i_N} a_{i_1,...,i_N}\sigma_1\otimes \sigma_2 \otimes \dots \otimes \sigma_N \equiv \sum_I a_I \sigma_I~,
\end{equation}
where we agglomerate the indexes referring to different qubits with $I$ to simplify the notation. We can define the difference of the state in a time step as
\begin{align}
	\ket{\Delta_U}=& \frac{1}{\Delta\tau}\left( \ket{\phi_m}-\ket{\phi_{m-1}} \right)= \frac{1}{\Delta\tau} \left(e^{-i \Delta \tau A_m} - 1 \right)\ket{\phi_{m-1}}~, \\
	\ket{\Delta_H}=& \frac{1}{\Delta\tau}\left(\frac{1}{c} e^{- \Delta \tau H} - 1 \right)\ket{\phi_{m-1}}~,\label{eq:diffITE}
\end{align}
where the first represents the difference from unitary evolution and the second the difference from ITE. The norm of their difference should be minimized, that is, $(\bra{\Delta_U}-\bra{\Delta_H})(\ket{\Delta_U}-\ket{\Delta_H})$, should be minimized. In order to obtain a solution, the unitary evolution is linearized for small time steps $\Delta \tau \to 0$, $\ket{\Delta_U} = -i A_m \Delta \ket{\phi_{m-1}}$. The product becomes
\begin{equation}\label{eq:difference_norm}
	\braket{\Delta_H|\Delta_H} + i \Braket{ \Delta_H | A_m |\phi_{m-1}} - i \Braket{ \phi_{m-1}| A_m |\Delta_H } +  \Braket{\phi_{m-1}|A_m^2 |\phi_{m-1}}~.
\end{equation}
If we write the matrix $A_m$ in terms of its decomposition in Pauli strings this expression becomes
\begin{equation}\label{eq:dif_min}
	f(\textbf{a}) = f_0 + \sum_I b_I a_I + \sum_{I,J} a_I S_{IJ} a_J~,
\end{equation}
where $f_0 = \braket{\Delta_H|\Delta_H}$, $b_I =i \Braket{\Delta_H|\sigma_I|\phi_{m-1}} - i \Braket{\phi_{m-1}|\sigma_I|\Delta_H}$, and $S_{IJ} = \Braket{\phi_{m-1}|\sigma_I \sigma_J|\phi_{m-1}}$. We can expand $b_I$ with Eq.~\eqref{eq:diffITE} so that it is expressed as an expectation value of operators in the state $\ket{\phi_{m-1}}$, after simplification it becomes
\begin{equation}\label{eq:bI_ex}
	b_I = \frac{i}{c\Delta\tau}\Braket{\phi_{m-1}|e^{-H \Delta \tau}\sigma_I-\sigma_I e^{-H \Delta \tau}|\phi_{m-1}}~.
\end{equation}
To minimize the difference between the ITE time step and the unitary one, we have to optimize \eqref{eq:dif_min} with respect to $a_K$, i.e., $\partial f(\textbf{a})/\partial a_k = 0$,
which sets the condition $b_K + \sum_J S_{KJ} a_J + \sum_I a_I S_{IK} = 0$, or in vector form $(S+S^T)\textbf{a} = -\textbf{b}$. Normally the matrix $S$ is singular, and a pseudo inverse or a least square solution of this system needs to be used.

\subsection{The algorithm}
With these ingredients in mind, we can detail the algorithm as follows. To evolve from state $\ket{\phi_{m-1}}$ to $\ket{\phi_m}$ we perform the following operations:
\begin{enumerate}
	\item Compute the matrix $S_{IJ} =\Braket{\phi_{m-1}|\sigma_I \sigma_J|\phi_{m-1}}$.
	\item Compute the norm $c=\sqrt{\Braket{\phi_{m-1}|e^{-2 \Delta \tau H}|\phi_{m-1}}}$.
	\item Compute $b_I = \frac{i}{c\Delta\tau}\Braket{\phi_{m-1}|e^{-H \Delta \tau}\sigma_I-\sigma_I e^{-H \Delta \tau}|\phi_{m-1}}$.
	\item Solve the linear equation $(S+S^T) \textbf{a} =- \textbf{b}$ by means of a pseudo inverse of matrix $(S+S^T)$. 
	\item Compute the Hermitian operator $A_m = \sum_I a_I \sigma_I$.
	\item Express unitary evolution $U=e^{-i A_m \Delta \tau}$ in terms of gates \cite{pauliunitary} if you are implementing the algorithm in a quantum computer, or in matrix form if you are doing a classical simulation.
	\item Finally evolve state as $\ket{\phi_m}=U\ket{\phi_{m-1}}$.
\end{enumerate}

\subsection{Trotterization for bigger systems}
So far, we have described how the imaginary time evolution of a Hamiltonian $H$ can be approximated by unitary operations. 
The problem, as with ITE, becomes increasingly costly as the size of the system increases. The number of expectation values needed to compute the matrix $S$ is $4^N$, which is exponential with the system size $N$. A Trotterization of the evolution step can be performed for ITE as
\begin{equation}
    e^{-H\Delta \tau}\approx \prod_k e^{-h[k]\Delta \tau}~,
\end{equation}
where we decomposed the Hamiltonian in $T$-local (act on $T$ consecutive qubits) pieces $H = \sum_k h_k$. The Trotterization has an error of order $O(\Delta \tau)$ and allows us to compute the easier exponentials $e^{-h[k] \Delta\tau}$, instead of the full Hamiltonian exponential. Likewise, we can perform a Trotterization of the unitary operations
\begin{equation}
    e^{-i A_m \Delta \tau} \approx \prod_k e^{-A_m[k]\Delta \tau}~,
\end{equation}
where $e^{-A_m[k]\Delta \tau}$ is the unitary evolution that approximates the imaginary evolution generated by the $h[k]$ piece of the Hamiltonia. Now, the previous algorithm can be repeated (at each time step) for each piece of the Hamiltonian $h_k$. Unfortunately, we again encounter the scaling problem if the $\sigma_I$ matrices are taken to act on all the qubits.
We can restrict ourselves to performing unitary evolution only around $D$ qubits, a domain centered around the qubits acted by $h_k$.
Here, the domain size $D$ refers to the number of qubits over which the approximate unitary operation acts during each local evolution step. A larger domain size allows capturing more entanglement and correlations, at the cost of increased computational resources.


\section{Variational Quantum Imaginary Time Evolution }
A variation of the previous algorithm called variational quantum imaginary time evolution (varQITE) was introduced in \cite{mcardle2019variational}. This new method efficiently simulates imaginary time evolution using hybrid quantum-classical computing. Since direct implementation \( e^{-\tau H} \) is unfeasible for large systems, a variational ansatz \( |\psi(\theta)\rangle \) is employed to approximate the evolved state
\begin{equation}
|\psi(\tau)\rangle \approx |\psi(\theta(\tau))\rangle~,
\end{equation}
where \( \theta(\tau) \) are some $\tau$-dependent parameters to be optimized. The variational form is usually represented using a parameterized quantum circuit
\begin{equation}
|\psi(\theta)\rangle = U(\theta) |0\rangle~,
\end{equation}
where \( U(\theta) \) is a unitary operation dependent on a set of tunable parameters \( \theta \). To obtain an equation of motion for \( \theta \), the McLachlan variational principle is employed,
\begin{equation}
\delta \| (H - E(\theta)) |\psi(\theta)\rangle \| = 0~,
\end{equation}
where \( E(\theta) = \langle \psi(\theta) | H | \psi(\theta) \rangle \) is the energy expectation value. This leads to a system of differential equations for the parameters
\begin{equation}
M_{ij} \dot{\theta}_j = -V_i~,
\end{equation}
with
\begin{align}
M_{ij} &= \text{Re} \left[ \frac{\partial \langle \psi[\theta] |}{\partial \theta_i} \frac{\partial | \psi[\theta] \rangle}{\partial \theta_j} + \frac{\partial \langle \psi[\theta] |}{\partial \theta_i} | \psi[\theta] \rangle \langle \psi[\theta] | \frac{\partial | \psi[\theta] \rangle}{\partial \theta_j} \right]
\end{align}
and 
\begin{align}
    V_i &= \text{Re} \left( \frac{\partial \langle \psi(\theta) |}{\partial \theta_i} H | \psi(\theta) \rangle \right)
\end{align}
where, "$\text{Re}$" is the real part of the argument. The matrix \( M \) and vector \( V \) are obtained after each iteration of the algorithm; the variational parameters \( \theta \) are iteratively updated using the update rule:
\begin{equation}
\theta_{t+1} = \theta_t - \eta M^{-1} V
\end{equation}
where \( \eta \) is the step size.

\subsection{Evaluating $M_{ij}$ and $V_i$ with Quantum Circuits}
To simulate imaginary time evolution using a variational quantum circuit, we need to evaluate two key quantities: the coefficient matrix $M$ and the vector $V$. These determine how the circuit parameters evolve over time. Each parameter in the circuit controls a gate $U_i(\theta_i)$. The derivative of such a gate with respect to its parameter can often be expressed as a combination of the gate itself and a simple Pauli operator (such as $X$, $Y$, or $Z$). Using this, the derivative of the full quantum state can be represented by modifying the original circuit slightly---inserting appropriate Pauli operators at specific locations. A widely used approach in the quantum computing community for estimating the gradient of expectation values of observables is the \textit{parameter-shift rule}. In this method, the gradient with respect to a given parameter is computed by evaluating the expectation value of the observable at two shifted values of that parameter, while keeping all others fixed. This technique is particularly relevant for computing the second term, denoted as vector $V_i$, in our formulation~\cite{Wierichs_2022}.
 The matrix element of $M$ and $V$ are then calculated as quantum overlaps: $M_{ij}$ involves the overlap between two modified circuits (i.e., derivatives with respect to $\theta_i$ and $\theta_j$), while $V_i$ involves the overlap between a modified circuit and the Hamiltonian applied to the original state.
These overlaps can be estimated on a quantum computer using interference-based circuits with an ancilla qubit \cite{mcardle2019variational}. The structure of these circuits allows efficient implementation using only a few extra gates.

\subsection{Ansatz}\label{sect:ansatz}
The variational part of the algorithm requires constructing an ansatz that can generate a general state dependent on some variational parameters $\theta$. 
The ansatz we considered consists of
a layer of rotations in $R_y (\theta)$ for each qubit, followed by a layer of rotations in $R_z (\theta)$, which contain the variational parameters. For a system of $N$ qubits, $2N$ variational parameters are needed. These variational gates are followed by a ladder of CNOT gates that generate entanglement between all the qubits. To make the algorithm more expressive, this ansatz can be repeated more times, with additional sets of variational parameters. Different strategies to construct the ansatz exist that can make the result more accurate. For instance, the ansatz can be constructed in such a way that respects the symmetries of the problem Hamiltonian, or different entangling strategies can be explored to improve accuracy. 

\section{Working Examples for the Traverse Field Ising Model}

We chose the transverse-field Ising model (TFIM) to showcase the effectiveness of each algorithm. This model has the advantage of being exactly solvable and displaying rich features, such as phase transition of the ground state, between a ferromagnetic and an antiferromagnetic phase. This model is described by the Hamiltonian
\begin{equation}\label{eq:TFIM}
     H = -J \sum_{i=1}^{N-1} Z_i Z_{i+1} + g \sum_{i=1}^{N} X_i~,
\end{equation}
where \( Z_i \) and \( X_i \) are the Pauli-\( Z \) and Pauli-\( X \) operators acting on qubit \( i \), \( J \) is the coupling constant, and \( g \) is the transverse field strength.
We restricted ourselves to a system of size $N=8$ qubits for demonstration purposes, and easy reproducibility with a desktop computer. We have been able to perform computations for systems of up to 20 qubits in HPC machines. 

For QITE, we divided this Hamiltonian into pieces that act on $T=2$ qubits, defined as
\begin{equation}\label{eq:trotTFIM}
    h[k]=JZ_kZ_{k+1} + \frac{g}{2}(X_k + X_{k+1})~,
\end{equation}
such that $H=\sum_k h[k]$. The matrices $A_m[k]$ are computed for different domain sizes, for $D=2$, which is the smallest domain size our pieces \eqref{eq:trotTFIM} admit, and for $D=4$ and $D=6$, which increasingly improve the fidelity of the evolution. The initial state was chosen to be the state $\ket{\uparrow}$ for all spins. 

For varQITE, we chose an ansatz that consists of two consecutive repetitions of the ansatz 
decribed in Section \ref{sect:ansatz},
which, for the example we chose ($N=8$ qubits), contains $32$ variational parameters. The initial state for varQITE is the state where all the variational parameters start with the value $\pi/3$. Which explains the difference in energy with the QITE counterpart that starts in the state $\psi(0)=\ket{0}^{\otimes N}$.

\begin{figure}
    \centering
    \includegraphics[width=0.49\linewidth]{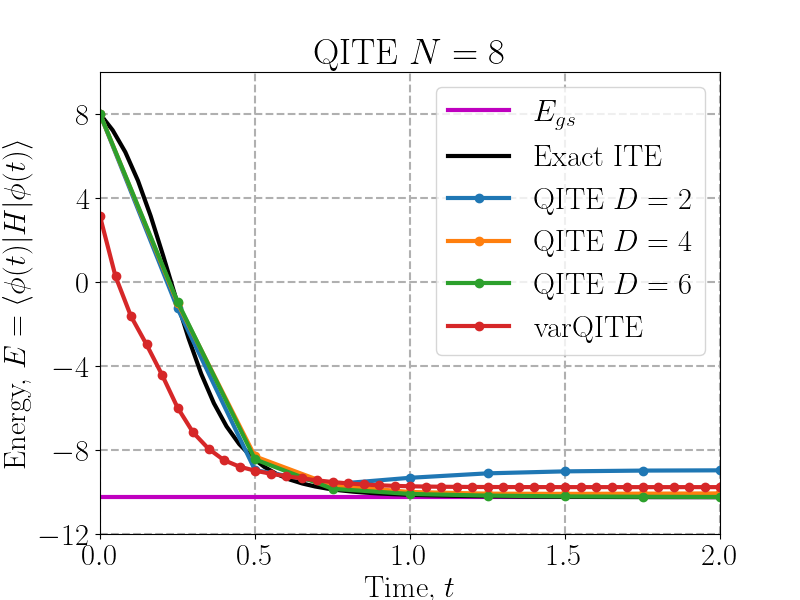}
    \includegraphics[width=0.49\linewidth]{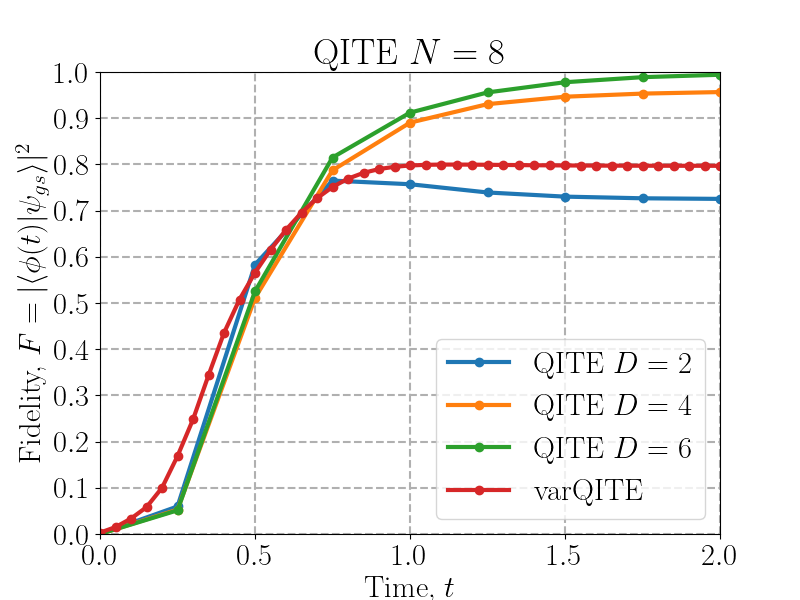}
    \caption{(Left) Energy of the state obtained with both algorithms and for different domain sizes $D$ of QITE. (Right) Fidelity between the exact ground state $\ket{\psi_{gs}}$ and the states obtained from the algorithms.
    The time step for QITE is $\Delta \tau=0.25$ and the time step for varQITE is $\Delta \tau=0.05$. The ground state $\ket{\psi_{gs}}$ and its energy $E_{gs}$  of the TFIM are calculated using Exact Digonalization.}
    \label{fig:energy}
\end{figure}

We compare in Fig.~\ref{fig:energy} the evolution of the expectation value of the energy and the fidelity with the exact ground state for both algorithms. In the left panel of Fig.~\ref{fig:energy} we compare these algorithms with exact Imaginary Time Evolution and see that convergence to the expectation value of the ground state is of the same time scale for all of them. In the right panel of Fig.~\ref{fig:energy}, we observe a similar trend for the fidelity of these methods with the exact ground state. 

\subsection{Discussion}

On the one hand, the accuracy of convergence in QITE depends on the size of the unitary domain \( D \), as expected: increasing \( D \) allows the algorithm to explore a larger subspace and improves the approximation of imaginary time evolution. On the other hand, the accuracy of varQITE does not depend on a single well-defined parameter like \( D \) in the case of QITE, but is instead strongly influenced by the choice of the variational ansatz and the initial configuration of its parameters. Different choices of entangling layers and parameter initializations can lead to widely varying outcomes, and there is no general prescription for designing an optimal ansatz. Variational algorithms, including varQITE, inherently involve a trade-off between \textit{expressivity} and \textit{trainability}. An ansatz that is too simple may lack the capacity to represent the ground state accurately, leading to convergence toward a local minimum rather than the global minimum. This issue is often confused with the barren plateau (BP) phenomenon but is conceptually distinct. The BP problem refers to the exponential suppression of both the gradient and its variance with system size, resulting in an almost entirely flat optimization landscape that hinders training from the beginning.
In contrast, when the optimization converges to a local minimum due to insufficient expressivity or poor initialization, it is not necessarily indicative of a BP. However, varQITE—like many variational algorithms—is still susceptible to the BP problem, particularly as the depth or expressivity of the ansatz increases. This imposes a practical limitation on scaling the method to larger system sizes. Balancing ansatz expressivity and trainability thus becomes a central challenge, and often must be approached heuristically, tailoring the ansatz structure to the problem at hand.

While QITE is a deterministic algorithm, that guarantees convergence to $\ket{\psi_{gs}}$ if condition \eqref{eq:overlap} is met, it is a costly algorithm to implement. The number of gates required to evolve one time step of QITE for $D=2$ is of the same order as two repetitions of the ansatz presented in 
Section \ref{sect:ansatz},
and, in our case, at least takes three steps are required to have convergence in energy, with a very bad performance in fidelity. For higher domains $D$, the number of gates is so big that, for current hardware, it is not possible to run it. Check Table 1 of \cite{pauliunitary} for quantitative values of the gate costs of a unitary step. Another advantage of varQITE come from the fact that it only requires computing the expectation value of the energy, while QITE requires order $4^D$ expectation values of Pauli strings in each time step, creating an overhead in circuit executions.  

\subsection{Python Code}

We provide two Jupyter notebooks that implement both algorithms for the Hamiltonian \eqref{eq:TFIM} that can be found at \url{https://github.com/dark-dryu/understanding-QITE-varQITE}. The QITE algorithm is coded with the sparse SciPy library, which allows to execute the algorithm for around 16 qubits on a desktop computer. The main runtime bottleneck of this algorithm comes from solving the linear system in step 4 of the algorithm. For a high $D$, it amounts to solving a linear system of $4^D$ equations, which is solved with a least squares routine. The variationl QITE algorithm is constructed from internal functions of the qiskit library \cite{qiskit2024} and allows user control of the ansatz through the \verb|EfficientSU2| function, which allows defining different single qubit gates, as well as various predefined entangling layers.

\section{Conclusions}

We have presented two algorithms that find the ground state of a problem Hamiltonian based on the imaginary-time evolution of the initial state, a non-unitary evolution, making direct implementation on quantum hardware unfeasible. The first algorithm, Quantum Imaginary Time Evolution (QITE), approximates imaginary time evolution using unitary operations, which can be implemented on a universal quantum computer. 
The second algorithm, varQITE, constructs a parametrized quantum ansatz of the state whose parameters are updated with a natural gradient descent, using energy as its cost function.
This approach is generally easier to implement on quantum hardware. However, it may suffer from the barren plateau (BP) problem, where the gradient vanishes exponentially with system size, making training intractable.
In our simulations, we do not observe signatures of the BP problem, but this does not guarantee convergence to the true ground state. The performance of varQITE is strongly dependent on the choice of ansatz. As mentioned previously, there is no systematic recipe for constructing an effective ansatz, and increasing its expressivity can exacerbate the BP issue. This challenge becomes especially pronounced for larger system sizes, where starting from a random initialization and lacking an informative ansatz construction strategy may render the algorithm impractical. There exists a fundamental trade-off between \textit{trainability}—the ability to optimize parameters effectively without encountering BPs—and \textit{expressivity}—the ability of the ansatz to represent the target state. This trade-off is prevalent in nearly all variational quantum algorithms. Therefore, in practical applications, one must heuristically balance these two aspects depending on the specific problem setting. In this paper, we applied both QITE and varQITE to the transverse-field Ising model and discussed the trade-offs and advantages that each method has.
A natural extension to this work, which would exploit the advantages of both algorithms, would consist of using a hybrid approach. First, QITE with a small domain could be used to get close to the ground state and use the resulting state as the initial ansatz of a variational routine, to avoid barren plateaus.
\begin{credits}
\subsubsection{\ackname} 
This work has been financially supported by the Ministry for Digital Transformation and of Civil Service of the Spanish Government through the QUANTUM ENIA project call- Quantum Spain project, and by the European Union through the Recovery, Transformation and Resilience Plan - NextGenerationEU within the framework of the Digital Spain 2026 Agenda.
T.P. acknowledges support from the Generalitat Valenciana under grant CIPROM/2022/66 for the research stay at the University of Valencia from January 12 to January 17, 2025. 
M.A.G.-M is supported by the European Union through the Recovery, Transformation and Resilience Plan - NextGenerationEU within the framework of the Digital Spain 2026 Agenda: also from Projects of MCIN with funding from European Union NextGenerationEU (PRTR-C17.I1) and by Generalitat Valenciana, with Ref. 20220883 (PerovsQuTe) and COMCUANTICA/007 (QuanTwin), and Red Tematica RED2022-134391-T. 
\subsubsection{\discintname}
The authors have no competing interests to declare that are relevant to the content of this article. 
\end{credits}

\bibliographystyle{unsrt} 

\bibliography{biblio}

\end{document}